\documentstyle[sprocl,epsfig]{article}

\def\be{\begin{equation}}
\def\ee{\end{equation}}
\def\bea{\begin{eqnarray}}
\def\eea{\end{eqnarray}}

\begin{document}
\vskip -1in
\rightline{SLAC-PUB-10663}
\rightline{August 19, 2004}
\rightline{}
\title{FAST DETECTOR SIMULATION USING LELAPS\footnote[1]{Work supported
in part by the Department of Energy contract DE-AC02-76SF00515.}\\}

\author{WILLY LANGEVELD}
\author{(Presented by Michael Peskin)}
\address{Stanford Linear Accelerator Center, Stanford, CA 94309}


\maketitle\abstracts{
Lelaps is a fast detector simulation program which reads StdHep generator files
and produces SIO or LCIO output files. It swims particles through detectors taking into
account magnetic fields, multiple scattering and dE/dx energy loss. It simulates
parameterized showers in EM and hadronic calorimeters and supports gamma conversions
and decays.\vskip .1in
\centerline{\it Presented at the International Conference On Linear Colliders (LCWS 04)}
\centerline{\it Paris, France, 19-24 April 2004}  
}

\section{Introduction}
Lelaps is a fast detector simulation program and a number of C++ class libraries,
the most important one being CEPack, the main simulation tool kit.
Main programs (lelapses) have been written for BaBar and for LCD (LDMar01, SDJan03
and SDMar04 are implemented). CEPack can also be used in conjunction with Geant4
parameterized volumes. In this way it is integrated in BaBar's Geant4-based
simulation. The standalone version for LCD reads StdHep generator files using the
(included) lStdHep class library, and produces SIO or LCIO output files that can
be read by JAS and LCDWired. It swims particles through detectors taking into
account magnetic fields, multiple scattering and dE/dx energy loss. It produces
parameterized showers in EM and hadronic calorimeters, converts gammas, supports
decays of certain short-lived particles (V decays), and it does all this very fast.

\section{CEPack}

The main class library is called CEPack containing the simulation
tool kit.

\subsection{Geometry}
Geometries are constructed using CENodes, which may contain a list of sub\-nodes.
A number of common CENode types are predefined (cylinders, cones, boxes, spheres etc.).
Transformations may be applied to CENodes in order to position and orient them.
Arbitrary affine transformations are allowed. CENodes need to provide methods for
determining distance along a straight line to entrance and exit, and for
determining whether points are inside them.

     \begin{figure}[t] 
     \begin{center}
     \vspace*{.2cm}
     \begin{tabular}{cc}
     \mbox{\epsfig{file=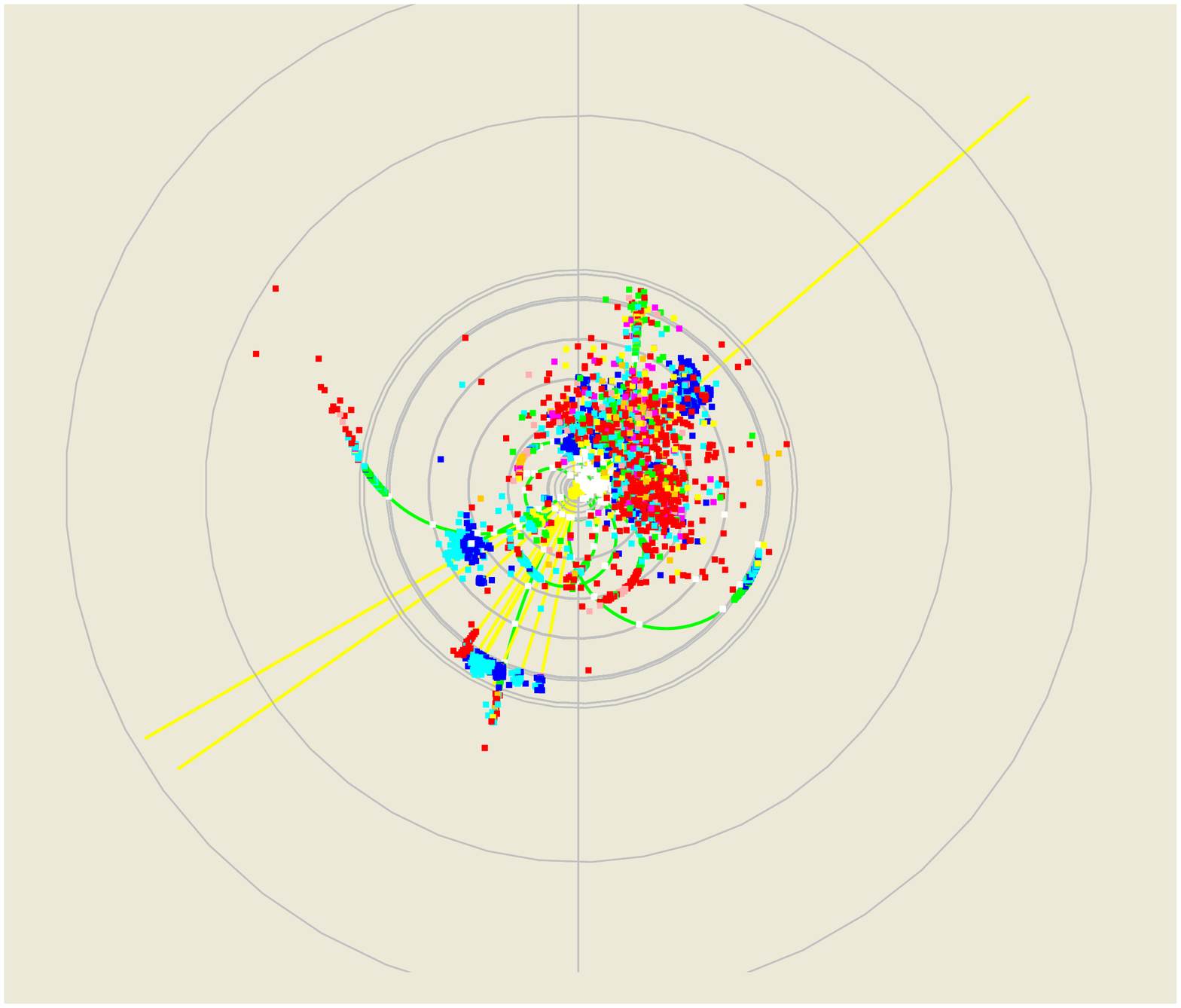,width=4.8cm,angle=270}}
     \mbox{\epsfig{file=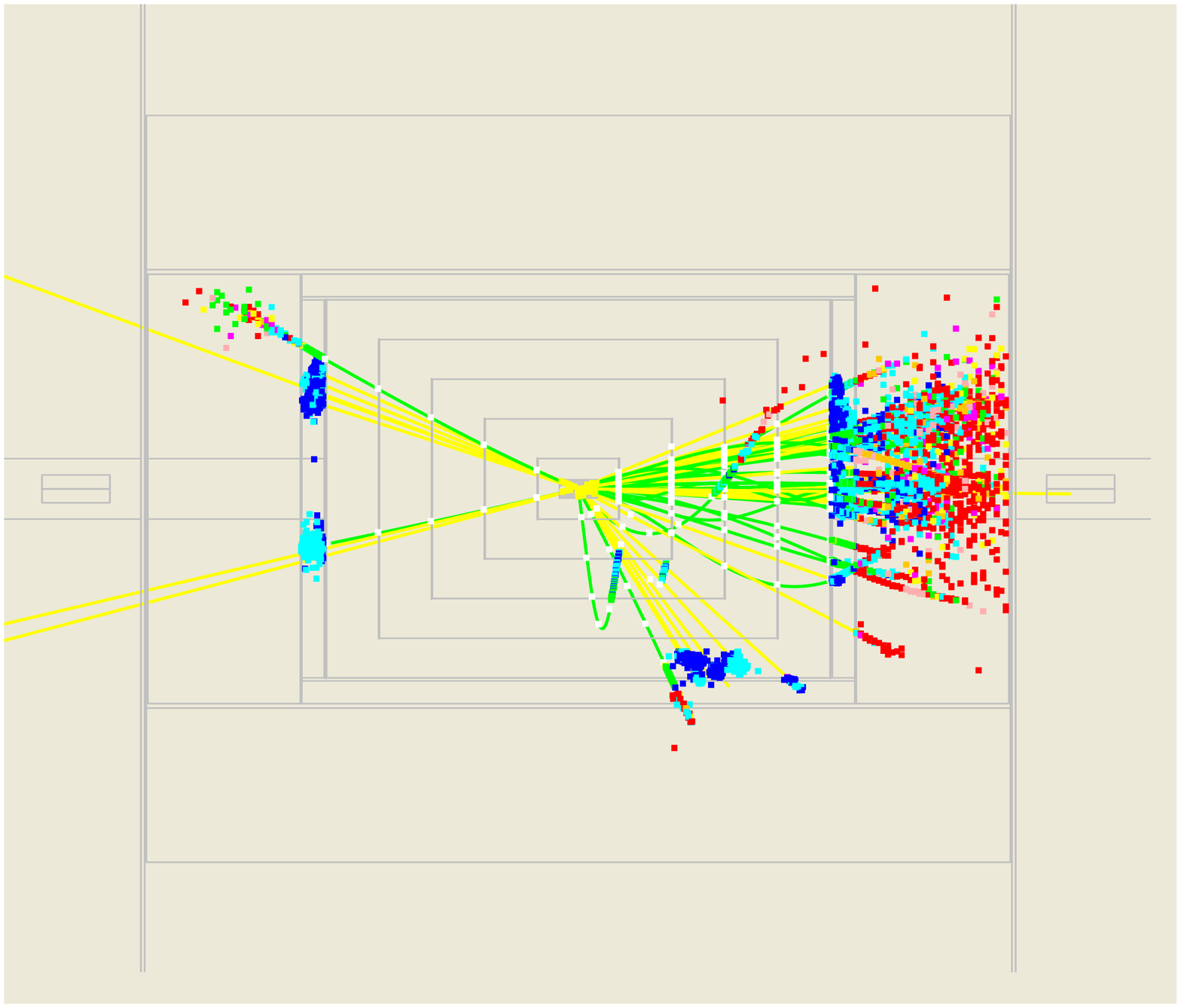,width=4.8cm,angle=270}}
     \end{tabular}
     \end{center}
     \caption{ZZ event in the SiD detector, as simulated by Lelaps.} 
     \label{fig:lelapssid}
     \end{figure}

CENodes may be assigned a numeric id and subid and may implement a method to
compute a subid from a location. This can be used to implement e.g. calorimeter
segmentation. CENodes do not have to be 3D objects: several predefined CENodes
consist of one or more 2D surfaces, which can be used, for exanple, to simulate
wire layers in drift chambers, or drift zones in TPCs. A Lelaps-simulated
ZZ event in the SiD detector is shown in figure \ref{fig:lelapssid}.

\subsection{Materials}
Specifying materials is very easy in Lelaps. All elements are built in with 
default pressure and temperature for gasses or density for solids and liquids.
Any compound can be specified by chemical formula and density or (for gasses)
temperature and pressure. Mixtures can be created by mixing elements and compounds
by volume or by weight. All needed quantities are calculated automatically. This
includes constants needed for multiple scattering and energy loss, radiation lengths,
interaction lengths and constants needed for shower parameterization.

\subsection{Matprop}
The Lelaps distribution comes with a little program called matprop that
allows one to view various material properties. An online version of
matprop is available at the URL

$$\rm http://www.slac.stanford.edu/comp/physics/matprop.html\quad.$$

\subsection{Tracking}
Tracking is performed by taking steps along a linear trajectory with endpoints
on a helix, such that the sagitta stays below a certain (settable) maximum.
CENodes have bounding spheres or cylinders; when computing distances
to CENodes, only relevant CENodes are considered. After each step, the amount
of material traversed is checked: if enough material was traversed, multiple
scattering and energy loss is performed and track parameters and the list of
relevant CENodes are updated. When an intersection occurs within a step, the
fractional step is executed, the CENode is entered, and the remaining fraction
of the step follows.

Multiple scattering is performed using the algorithm of Lynch and Dahl[1].
Material is saved up along the track until there is enough. dE/dx is calculated
using the methods by Sternheimer and Peierls[2]. All constants are precalculated
by the material classes.

\subsection{Shower parameterization}
Electromagnetic showers are parameterized using the algorithms of Grindhammer and Peters[3].
Calorimeters are treated as homegeneous media. The longitudinal shower profile is given
by a gamma distribution with coefficients depending on the material (Z) and energy.
The profiles are fluctuated and correlations between the coefficients are taken into account.

For each step of one radiation length, a radial profile is computed consisting of
two distributions, one describing the core of the shower and the other the tail.
Various parameters are functions of Z, shower depth t and energy. The energy to be
deposited is divided into spots thrown in radius according to the radial profile,
and uniformly in $\phi$ and between t and t+1. Roughly, about 400 spots are generated
per GeV of shower energy and reported as hits.

Hadronic showers are parameterized in a similar way, with some modifications.
The location where the shower starts is simulated using an exponential law with
attenuation given by the interaction length. The longitudunal profile uses the Bock
parameterization[4]. A combination of two gamma distributions, one using radiation
lengths and the other interaction lengths, is used. The Bock parameterization does
not specify radial profiles. For the moment we use a radial profile similar to
Grindhammer and Peters (for EM showers) but with radiation lengths replaced by
interaction lengths and faster spread with depth. The parameters still need to
be fine-tuned.

These parameterizations were compared to results from Geant4[5].
In general pretty good agreement was found for EM showers. Hadronic showers agree
pretty well longitudinally, but not as well radially. Hadronic shower parameterization
has been tweaked since then, but needs further work.

\subsection{Decays and gamma conversions}
CEPack supports decays of unstable particles and gamma conversions. Supported unstable
particles are $\pi^0$, $K^0_s$, $\Lambda$, $\Sigma^{+/-/0}$,
$\Xi^{-,0}$ and $\Omega^-$. Only decay modes with branching fractions greater than 2\%
are supported (mostly ``V decays'').

\section{Using CEPack in Geant4}
To use CEPack inside Geant4 one subclasses G4VFastSimulationModel. In its setup()
method, one creates the CENode and subnodes corresponding to the CEPack geometry.
In the Doit() method one converts from G4FastTracks to CETracks and
calls the CETrack's swim() method. By subclassing CETrack, all hits are reported
using the CETRack report\_hit() method. One then converts hits to one's favorite format
and updates Geant4's notion of the track using G4FastStep (or calls KillPrimaryTrack()
if the track has ended).

\section{CEPack and Lelaps}
Lelaps for LCD is a standalone program which sets up the CEPack
geometry, reads input files and produces output files. Currently supported
geometries are LDMar01, SDJan03 and SDMar04. To read generator level event files in StdHep
format, it uses class lStdHep (``StdHep light'', included in the distribution).
StdHep particles are converted to CETracks and tracked through the geometry.
When hits are reported, they are added to SIO or LCIO hit lists. For calorimeter hits,
the spots are first accumulated and turned into energy depositions in individual
calorimeter cells and then added as hits. Finally, the SIO or LCIO event structure
is written out. 

\section{Performance}
Taking the example of a typical $e^+ e^- \rightarrow Z Z$ input file, the performance of
Lelaps is as follows. With tracking alone, Lelaps can simulate 3--4 events per second
(at 1 GHz processor speed) for the LD detector, or about 2 events/s for the SiD detector.
Adding parameterized showering costs 15\% (SiD) to 30\% (LD). Adding decays and conversions
takes another 20\%. Writing an LCIO output file (14 MB compressed for 100 events) costs
another 40\%. SIO output takes much longer (a factor 2 to 4 depending heavily on
calorimeter segmentation)---this could be optimized, but SIO is a deprecated format.

The performance is slightly platform/machine dependent. On a Linux machine,
tracking takes 0.281 seconds/event, on a Solaris machine 0.154
seconds/event, and on Windows (cygwin with gcc 3.2) 0.384 seconds/event (all at
1 GHz processor speed).

\section{Future}
The Lelaps\footnote[2]{Lelaps (storm wind) was a dog with such speed that, once set upon a chase,
he could not fail to catch his prey. Having forged him from bronze, Hephaestus gave him to
Zeus, who in turn gave him to Athena, the goddess of the hunt. Athena gave Lelaps as a wedding
present to Procris, daughter of Thespius, and the new bride of famous hunter Cephalus.
A time came when a fox created havoc for the shepherds in Thebes. The fox had the divine
property that its speed was so great that it could not be caught. Procris sent Lelaps to
catch the fox. But because both were divine creatures, a stalemate ensued, upon which Zeus
turned both into stone. Feeling remorse, Zeus elevated Lelaps to the skies, where he now
shines as the constellation Canis Major, with Sirius as the main star.} and CEPack interfaces are not yet frozen. Changes may be necessary
considering the planned new features and improvements. Some of these are:
support for combinatorial geometry; allowing shower continuation into a next
volume and reading geometry descriptions from a standard file format.
Also, hadronic showers need further tuning. 

\end{document}